# Non-Superconducting Non-Abelian Statistics in One-Dimensional Topological Insulators

Zhigang Song

[1]Department of Engineering, University of Cambridge, JJ Thomson Avenue, CB3 0FA Cambridge, U.K.

[2]State Key Laboratory for Mesoscopic Physics and School of Physics, Peking University, Beijing 100871, P.R. China

## Abstract

Topological materials are of great interest for applications in quantum computing, providing their intrinsic robustness against the environmental noises. One popular direction is to look for Majorana modes in the integrated systems interfaced with superconducting materials. However, is the superconductivity necessary for the materials to exhibits non-abelian statistics? Here we predict with strong theoretical and numerical evidences that there exist topologically phases in a class of one-dimensional single crystals, which exhibit large bandgaps and are within experimental reach. Specifically, the nontrivial Zak phases are associated with gapless boundary states, which provide the non-Abelian statistics required for constructing topologically-protected quantum gates, even without superconductivity and magnetic field. Another anomalous feature of these materials is that their electric polarization is quantized and a transverse field can induce a longitudinal polarization. This novel physical property provides an experimentally-accessible mechanism for braiding the non-Abelian anyons.

**Introduction**

Quantum computation and simulation are promising methods for solving computational and physical problems that are not tractable with current classical computers[1]. In reality, the major obstacle in building a practical quantum computer comes from the interaction with the environment, which causes decoherence to the qubits. To solve such a problem, a fault-tolerant architecture of quantum computation requires a repeating implementation of quantum error correction, which consumes many more extra physical qubits to represent a logical qubit. Alternatively, one may exploit the physics of topological materials to protect against noise, an approach known as topological quantum computing (TQC) [2]. Standard implementation of TQC is based on either zero-energy Majorana modes or anyons in fractional quantum Hall liquids[3,4]. However, the fractional quantum Hall effect can only be observed under high magnetic fields and at ultra-low temperatures. Furthermore, the existence of a physically robust method in manipulation of the anyons, known as braiding, remains unclear. On the other hand, Majonara modes are predicted to exist in the heterostructures of semiconductors or topological insulators coupled to superconductors[5-7]. Previously, large spin-orbit coupling (SOC), Zeeman splitting, and superconducting proximity interactions are necessary ingredients for realizing the Majorana modes. Currently, it remains a major experimental challenge to observe the non-Ablien phases of the Majorana modes in condensed-matter systems, although some groups have reported indirect evidences on the existence of Majorana modes[8,9]. Overall, the experimental challenge for realizing topological quantum computation would be significantly reduced, if one can realize non-Abelian anyons in a pristine material without extreme conditions (e.g. low temperature for maintaining superconductivity and high magnetic field). This motivates our search for non-Abelian anyons based on one-dimensional topological insulators.

Similar to one-dimensional topological superconductivity, a topologically-protected bound states may potentially exist in one-dimensional topological insulators due to band inversion[10,11]. When a Bloch particle is evolved adiabatically through a one-dimensional parameter space, the acquired Berry phase is known as Zak phase[12]. Generally, the Zak phase returns to its original value after an adiabatic evolution in a cyclic trajectory in parameter space; otherwise a nonzero nontrivial value is generated. The manifestation of a quantized change of the Zak phase leads to observable physical consequences in optical lattices[13], cold atoms[14], and acoustic systems[15]. Furthermore, a variation of Zak phase enables the realization of fermion-number fractionalization[16], spin-charge separation[17], Wannier-Stark ladders[18]. Based on the use of the Zak phase, the concept of one-dimensional topological insulators has been proposed in the SSH model[16], and some efforts have been poured to proposing possible systems, such as modulated quantum wires, quantum dots and double atom ladders under a periodic magnetic fields[19,20]. A pristine crystal with the research of experiments that is topologically nontrivial has not been found.

Here we show that materials with band inversion and SOC can result in a boundary state that is analogous to a Majorana mode, which can be applied to topological quantum computation. Specifically, we present a series of single-crystal materials with nontrivial Zak phases, where the boundary states behave effectively as non-Abelian anyons. We further show that even in the absence of a strong electronic interaction, a universal TQC can still be realized with these non-superconducting materials, where the qubit size is only of order of a few nanometers. In addition, it is possible to braid the boundary anyons with purely electronic methods, which is due to the existence of a novel mode of electronic polarization in these materials. Finally, we show that the bandgap of these topological materials can be very large, which raises the possibility where topological qubits may be protected even at the room temperatures.

**Results and Discussions**

**Band structure and boundary states**

In the following, we shall focus on the $CuO_2$ chain as a concrete example to construct our topological model; the main conclusions are also applicable to other 1D materials, such as $TiI_3$, $CuBr_2$, $Ta_4SiTe_4$ and $Nb_4SiTe_4$ chains. The optimized structure of a free-standing linear $CuO_2$ chain is shown in Fig. 1(a). The equilibrium lattice constant is $a$=2.72 Å. The neighboring Cu atoms are bridged through two O atoms, and a Cu atom sits at the center of a rectangle consisting of four O atoms (see Fig. 1(a)). The phonon calculation of linear $CuO_2$ chain exhibits no imaginary frequencies, indicating that the one-dimensional $CuO_2$ is locally stable. Three-dimensional $CuO_2$ is stacked by linear $CuO_2$ chains with staggering of half a period, and the bulk crystal structure is shown in Fig. S1. Besides, $TiI_3$, $Ta_4SiTe_4$ and $Nb_4SiTe_4$ bulks are stacked with corresponding chains by a weak van Waals interaction. The chain and bulks of $TiI_3$, $CuBr_2$, $Ta_4SiTe_4$ and $Nb_4SiTe_4$ chains are shown in the Figs. S1(c-g).

The ground state of one-dimensional $CuO_2$ is computed to favor ferromagnetism with a magnetic moment of 1 $\mu_B$ per unit cell. The magnetic moment of ferromagnetic state is perpendicular to plane $M_3$, and thus the initial mirrors $M_1$ and $M_2$ of atom sites are broken (see

Fig. 1(a)). The ferromagnetic state is 134 meV per cell lower than possible antiferromagnetic configuration of nearest staggered spins. These results are in good accordance with a previous work, which claimed an antiferromagnetic exchange coupling between the chains and a ferromagnetic ordering within the chain[21]. Fig. 2 (a) implies that the majority-spin and minority-spin bands are split by the exchange interaction, and only the majority-spin bands are in the vicinity of the Fermi level. The 3$d$ shell of Cu is partially filled, leaving 1 unfilled $d$-orbital of the majority spin. The majority spin bands cross the Fermi level at two distinct points, forming a pair of gapless Dirac cones with a linear dispersion. The two Dirac points are degenerated due to the inversion symmetry. A maximum bandgap of 20 meV is opened when SOC is included in the calculations, and the conduction and valence bands become well-separated (see Fig. 2(b)).

Due to a small magnetocrystalline anisotropy energy of approximately 0.04 meV per cell, the magnetic moment can be easily orientated by an external magnetic field, but the ferromagnetism remains invariant. When the magnetic moment is perpendicular to the $CuO_2$ plane, the bandgap is up to 20 meV, but when the magnetic moment is in plane, the bandgap vanishes. As the bandgap is dependent of the orientation of magnetic moment, the linear $CuO_2$ chains can serve as a spin-filtered electronic switch, controllable through an external magnetic field. The on/off state can be realized by rotating the external magnetic field. In our DFT calculations, the magnetic moment relaxes to a configuration of (0.57, 0.58, 0.57) $\mu_B$, which is the configuration taken for the following calculations. The fundamental bandgap is approximately 14 meV for this magnetic configuration.

One of the key features of our model is a novel phase-transition mechanism induced by a transverse electric field. Two orthogonal electric fields vertical to linear $CuO_2$ chains are applied (shown in Fig. 1(a)). The electric field $E_2$ has indiscernible influences on the band structure. However, the inversion symmetry can be broken by electric field $E_1$ which induces an unbalanced potential distribution (see Fig. 1(b)). Let us denote gap1 and gap2 to represent the local gaps at the two Dirac points, which are labeled as K and K', respectively. It can be seen from Fig. 2 (c) that the gap1 increases linearly with the electric field $E_1$, while the gap2 decreases to zero at first, then reopen, and then increases further with increasing $E_1$. The closing and reopening of the bandgap at a critical electric field of $E_1$=0.08 V/Å corresponds to a process of band inversion accompanied with a phase transition.

The effects of SOC and electric fields are manifested in the electric Zak phase acquired by an adiabatic motion over the Bloch bands. The electric Zak phase is obtained in a discrete Brillouin zone by $\gamma \approx -\sum_{n}^{N_k-1} \text{Arg}(\det \Psi_n^\dagger \Psi_{n+1})$, where $\Psi_n$ is composed of the valence eigenstates as column wave vectors at $k = n2\pi/N_k$. The Berry connection and electric Zak phase are gauge dependent, while the winding number $\mathbb{N} = \frac{1}{\pi}\big(\gamma(\text{R}_f)-\gamma(\text{R}_i)\big) \bmod 2$ is gauge invariant. A change of winding number signifies a topological phase transition, which can be applied to distinguish a nontrivial or trivial phase by calculating electric Zak phase variation before and after including SOC.

Our results indicate that the winding number is 1 in the free standing linear $CuO_2$ chain after including SOC, indicating that the ground state is nontrivial. Under the external electric fields $E_1$, the winding number switches from 1 to 0 when the external electric fields $E_1$ sweeps across a critical electric field of $E_1$=0.08 eV/Å. The alternation of winding number is in good accordance with the gap closure and confirms the phase transition above. In one dimension, the electric polarization is proportional to the electronic Zak phase if there is no structural distortion. Since our system holds under the inversion symmetry, the change of the electric Zak phase can exactly reveal the information about the electric polarization. The individually-calculated electric polarization and the electric Zak phase agree with each other (see Fig. 2(b)). During the phase transition, the Wannier center of bulk cell shifts by $\frac{a}{2}$, resulting in a quantum polarization $\frac{ae\gamma}{2\pi}$, and half a charge is transferred from bulk to each edge in a finite nanowire. Transverse electric susceptibility becomes infinitely large near the phase-transition point owing to the quantized variation of the Zak phase. Note that the connection between the Zak phase and the bulk polarization also exists in the Su-Schrieffer–Heeger and the Rice-Mele model[22]. However, the key difference is that in our model, the response is not along the same direction; a longitudinal electric polarization can be induced by a transverse electric field. Furthermore, the transferred charge here is fractional. This type of materials can work as a reversible toggle of logical storage devices or switches.

Specifically, the electric polarization is proportional to Zak phase as follow,

$$p = \frac{ae\gamma}{2\pi} + nae. \tag{1}$$

where $e$ is the elementary charge, and $a$ is lattice constant. Eq. 1 implies that fractionalization of transferred charge and quantized electric polarization are two distinct features of the one-dimensional materials with a nontrivial Zak phase. The fractionally-charged quasiparticles can be detected by a quantized current; a similar method has been proposed in a previous report[23].

**Microscopic mechanism**

To investigate the microscopic mechanism of the phase transition and nontrivial topology, we constructed a tight-binding model using the $k \cdot p$ theory subject to the symmetry constraints. The on-site SOC is omitted because of the initial orbitals in the vicinity of the Fermi level. The angular-momentum-resolved calculations also show that the onsite SOC is very small. The nearest, next nearest neighbor hoping, and nearest neighbor SOC are included in the Hamiltonian

$$H = \sum_j \varepsilon_j c_j^+ c_j + \sum_{<jk>} -t c_k^+ c_j + i t_{soc} v_{jk} c_j^+ \sigma_z c_k + h.c. + \sum_{<<jk>>} -t' c_j^+ c_k + h.c., \quad (2)$$

where $\varepsilon_j$ is onsite potential, and $c_j^+$ ($c_j$) denotes the fermionic creation (annihilation) operator for spin polarized electrons. $t'$, $t$ and $t_{soc}$ are hoping amplitudes. This model is distinct from the SSH model by a single-spin Dirac cone and SOC. After performing a Fourier transformations followed by a canonical transformation (details are in supplementary information), we obtain a continuum model in two dimension as follows,

$$H = \begin{pmatrix} \tilde{\varepsilon}_d + 2\tilde{t}_d \cos(k_y y) & i(4t_{soc} \cos(k_y y/2) + \alpha t_{pd} \sin(k_y y/2)) \\ -i(4t_{soc} \cos(k_y y/2) + \alpha t_{pd} \sin(k_y y/2)) & \tilde{\varepsilon}_p + 2t_p \cos(k_y y) \end{pmatrix} \quad (3)$$

where the parameters of $\tilde{\varepsilon}_d = -0.562$ eV and $\tilde{\varepsilon}_p = 0.087$ eV are effective onsite potentials for $d$ and $p$ orbitals, $t_p = 0.12$ eV ($\tilde{t}_d = -0.599$ eV) is effective hopping amplitude of the nearest $d$ ($p$) orbitals, and $t_{soc} = 0.006$ eV is SOC interaction. The external electric field and hopping between neighboring $d$ and $p$ orbitals are embedded in the parameter $\alpha$. Although both SOC and electric field can open the bandgaps at the Dirac points, the electric Zak phase with SOC alone is always distinct from the electric Zak phase with only a large electric field by a value of $\pi$. Therefore, the phases of the linear $CuO_2$ chains are distinct when considering SOC and an electric field alone. In the vicinity of the two valleys, the band gap is determined by $\left| 2t_{soc} + \frac{\sqrt{3}}{2} \alpha t_{pd} \right|$. A band crossing is inevitable between the two phases during an adiabatic transition. Actually, the system with a gap opened by SOC alone is topologically nontrivial. However, the bandgap can remain open during the progress of a sweeping electric field, if there is a perturbation in the format of $U\tau_x$, where $U$ is real, and $\tau_i$ is a Pauli matrix on the same basic set of the Hamiltonian. Some symmetries do not permit the perturbations in the format of $U\tau_x$, such as inversion symmetry or the product of the mirror reflection $M_1$ and time-reversal symmetry, $\tilde{M}_1 = -i\sigma_y CM_1$ (spinless mirror), where $C$ is the complex conjugation operation. $\tilde{M}_1 H \tilde{M}^{-1} = H$ and $\tilde{M}_1 U\tau_x \tilde{M}^{-1} = -U\tau_x$, and $IHI^{-1} = H$ and $IU\tau_x I^{-1} = -U\tau_x$. If an electronic system maintains such a spinless mirror symmetry or inversion symmetry, the perturbations in the format of $U\tau_x$ is forbidden. Actually, the Hamiltonian in Eq. 3 describes a topological crystal insulator protected by spinless mirror symmetry or inversion symmetry. Thus, the phase transition aforementioned under the electric field is indeed a topological phase transition. Although spinless mirror symmetry or inversion symmetry is necessary for the topology, the variation of Zak phase $\pi$, namely quantum polarization, can be robust only with $C_{2x}$ symmetry according to our further DFT calculation, when we remove the spinless mirror symmetry and inversion symmetry.

To investigate the relationship between the transferred charge and edge states, we use different methods to calculate the density of states and energy levels of the end states. Fig. 3(a) shows the energy levels obtained by solving Eq. 3 in a finite but long enough strip. There is a pair of degenerate energy levels residing in the bulk gap. According to the number of valence electrons, only either of the two edge energy levels is occupied, and the Fermi level is across the middle of the two energy levels. Both degenerate energy levels are equally projected on both edges of the strip (see Fig. 3(b)). The total transferred charge to each edge is 0.5e, which is well consistent with the bulk electric Zak phase. Combing the recursive Green-function method and the tight-binding Hamiltonian in Eq. 3, we obtained the density of states at the edge of the semi-infinite geometry (Fig. 3(a)). When a large electric field is applied, the density of states consists of two peaks. In the contrast, the two peaks degenerate to one, if the electric field is small. The edge density of states is in good accordance with the calculated energy levels. According to our numerical calculations, the wave functions of the highest occupied state and the lowest unoccupied state are orthogonal. Both the electrons of the highest occupied state and the lowest unoccupied state distribute equally at the two ends. Thus the half-charged quasiparticle is robust despite of the degeneracy of the highest occupied state and the lowest unoccupied state. As discussed above, there are two topologically-distinct phases characterized by the bulk electric Zak phase. The topological phase transition is accompanied by a Zak phase variation of $\pi$.

We show two typical types of edge states of nontrivial phase in Figs. 4(b, c). A soliton forms at an edge or the domain wall of nontrivial and trivial zones. The soliton is half charged according to integration of charge distribution, and its antiparticle is itself. The edge states cannot be removed by a shift of the cropped tails, which is in accordance to the fact that the winding number cannot be removed by a gauge transformation. The degenerate end states of nontrivial phase are thus topologically protected and independent of the details of the edges. In a neutral and finite system, a soliton with half of an electron resides in one edge, and a soliton with half of a hole is located in the other edge, leading to a dipole despite of an inversion center in bulk unit cell.

The topological continuum model and the idea of Zak phase are general, although we discuss the physics based on the certain material and symmetry. The nontrivial phase is characterized by the band inversion[24] and SOC, which is similar to two-dimensional quantum spin Hall[25] and three-dimensional topological matters. It is naturally distinct from the polyacetylene, in which the boundary state can only exist in its interface of two distorted phases rather its boundaries[26]. By data mining, $TiI_3$[27,28], $Ta_4SiTe_4$[29] and $Nb_4SiTe_4$[30] chains are found to be such topologically nontrivial materials with inversed bands and large enough SOC. The calculated band structures and the boundary states are shown in the Fig. 4. It is necessary to state that the boundary states with a half quantum number additionally require a broken time-reversal symmetry. According to our calculation, $Ta_4SiTe_4$ bulk and $Ta_4SiTe_4$ chains are time-reversal-symmetric. Although they have zero Zak phase due to time reversal symmetry, the band inversion and SOC coupling also lead to a topologically-protected two-fold boundary states. The edge state possesses integer transferred charges or charged with integer electrons (or hole). The details are shown the supplementary in formation.

Although free standing linear $CuO_2$, $TiI_3$, $CuBr_2$, $Ta_4SiTe_4$ and $Nb_4SiTe_4$ chains have not been synthesized, they are easy to exfoliate from their bulk due to the weak van Waals inter-chain interaction. In our lab, we have even synthesized $Ta_4SiTe_4$ chains (details seen in the Part Ⅳ of the Supporting Information). The calculated intra-chain magnetic coupling is only 4 meV in the $TiI_3$ chains, and a small magnetic field or magnetic substrate is necessary to stabilize the magnetism of the $TiI_3$ chains. Linear $CuO_2$ chains exist in a lot of bulk materials with a variety of intercalated ions (Ca, Sr, Ba, H, La, Y), such as $(Sr_2Cu_2O_3)_xCuO_2$[31], $Ca_xCuO_2$[32], $Li_xCuO_2$[33], $Cu(OH)_2$[34]. The stoichiometry and elementary species of the intercalated ions vary in a large range, so a three-dimensional $CuO_2$ stacked by linear $CuO_2$ chains without intercalated ions is thought as a stable crystal. the inorganic crystal structure database (ICSD) with the assigned numbers of 150866, 89236 or 150913[35]. More recently, single chains of $MnO_2$, $FeO_2$, $CoO_2$ and $NiO_2$ are experimentally synthesized on the Ir surfaces[36].

**Non-Abelian statistics and quantum computing**

In a charge-neutral system, the left ($\psi_L$) and right ($\psi_R$) boundary states can be rewritten as:

$$\psi_\pm = \psi_L \pm e^{i\alpha}\psi_R \qquad (4)$$

where $\alpha$ is an arbitrary phase that can be tuned by temporarily overlapping $\psi_L$ and $\psi_R$, and then separating them. The boundary states are in the following form, $\psi_{L/R} = \beta e^{-\left|\frac{y-y_{L/R}}{\xi}\right|}$, where $\xi$ is proportional to the bandgap, and thus the bandgap is easy to be tuned by nonlocal operations such as electric fields.

The creation and annihilation operations corresponding to $\psi_\pm$ are given by

$$\Gamma_\pm = \int dr\, \psi_\pm \varphi = \frac{1}{\sqrt{2}}(\Gamma_L \pm e^{i\alpha}\Gamma_R) \qquad (5)$$

where $\Gamma^\dagger_{L/R}$ and $\Gamma_{L/R}$ are creation and annihilation operations on the left and right states, respectively. The signs of $\pm$ in the Eqs. 4-5 indicate a chiral degree of freedom. Now we define the parity as

$$P_\pm = 1 - 2\Gamma^\dagger_\pm \Gamma_\pm \qquad (6)$$

which has eigenvalue of $\pm 1$, and the parity is conserved.[37] In the nonmagnetic nanowires of Ta$_4$SiTe$_4$ and Nb$_4$SiT$_4$, the constant of 2 in the Eq. 6 is neglected since the boundary charges are integers in unite of elementary charge. After the quasiparticle on the end states are exchanged, we have

$$U^\dagger_{LR}\Gamma_L U_{LR} = \Gamma_R,\; U^\dagger_{LR}\Gamma_R U_{LR} = e^{-2i\alpha}\Gamma_L,\; U^\dagger_{LR}\Gamma_\pm U_{LR} = \pm e^{-2i\alpha}\Gamma_\pm \qquad (7)$$

It is obvious that the unitary operator

$$U_{LR} = e^{-i\alpha(\Gamma^\dagger_+\Gamma_+ + \Gamma^\dagger_-\Gamma_-)+i\pi\Gamma^\dagger_-\Gamma_-} = e^{-i(\alpha-\frac{\pi}{2})(\Gamma^\dagger_L\Gamma_L + \Gamma^\dagger_R\Gamma_R) - i\frac{\pi}{2}(e^{i\alpha}\Gamma^\dagger_L\Gamma_R + h.c.)} \qquad (8)$$

is the desired braiding operator. Note that $[U_{mn}, U_{nk}] \neq 0$, where $m, n, k$ denotes different boundary states, and thus the boundary quasiparticles obey a non-Abelian statistics. Similar methods to construct non-Abelian anyons have used for system in periodic magnetic field in the previous reports[19]. Mathematically, the braiding operations are a representation of the multiple-strand braid group. Standard methods of encoding by Majorana modes can be applied here[38-40].

However, in contrast with the Majorana-based qubits, the topological qubits here can be charged, and thus they can be manipulated and readout by well-developed electronic methods. In particular, if we apply a small potential difference across the two ends, the charge density mainly becomes localized at only one end. If the potential difference is reversed, the charge density becomes localized at the opposite end. The charge distribution under a small electric potential difference, which is equal to 0.1% of bandgap, is plotted in Fig. 3(a). Since the boundary states are sensitive to the potential gradient (or equivalently an electric field), the boundary states are possible to braid by electronic methods. Similar methods have been widely applied in quantum-dot qubits. Here, we take a Y-junction as an example (see Fig. S5(b) in the Supporting Information), the braiding progression is shown in Figs. S6(c-f) in the part III of Supplementary information.[40]

As usual, one requires four anyons, formed by two fermions on the nanowires, to encode a qubit. The Hilbert space is spanned by the eigenvalues of the fermion number operation $|n_1, n_2\rangle$, and the first nanowire specifies a choice of basis states for logical bits. In the two-dimensional Hilbert space with an odd parity, $|0\rangle \to |01\rangle$ and $|1\rangle \to |10\rangle$, and the braiding operation matrixes are as follows:

$$U_{12} = \begin{pmatrix} 1 & 0 \\ 0 & e^{-i\alpha} \end{pmatrix}, U_{34} = \begin{pmatrix} e^{-i\alpha_3} & 0 \\ 0 & 1 \end{pmatrix} \quad (9)$$

where the paremeters $\alpha_1, \alpha_3$ are set as $\pi/4$, $\pi$, respectively. By setting $\alpha_3 = \pi/2$

$$U_{23} = \begin{pmatrix} 0 & 1 \\ 1 & 0 \end{pmatrix} \quad (10)$$

Thus the π/8 phase, Z and T gates are given by

$$T(\frac{\pi}{8}) = e^{i\frac{\pi}{8}} U_{12}, Z = -U_{34}, X = U_{23} \quad (11)$$

Details are shown in the part III of the supplementary information. The operations of T gate with π/8-phase, In Eq.11, which is missing in the Majorana-based TQC. Furthermore, six anyons are sufficient to form two qubits. The physical basis is specified in the even-parity subspace $|Q_1, Q_2\rangle = |n_1, n_2, n_3\rangle$, and we label logical bits by the second and third nanowires. Thus, the controlled-NOT gate is as

$$\text{CNOT} = U_{56} \quad (12)$$

Since two noncommutable single-bit operations (here NOT and T gates), and a two-qubit CNOT gate constitute a universal gate sets, the topological quantum computation based on such anyons is universal[41,42]. Moreover, it is possible to scale up the qubit size by a honeycomb structure.

Finally, let us discuss the fault tolerance of the qubits. The decoherence of the qubits comes from the interactions with the environment. As the energy levels are degenerate, the dynamic phase remains an unphysical global phase. The braiding operation is non-Abellian and discrete, so the phase is robust. Furthermore, qubits are encoded in the ground states, which are stable at sufficiently low temperatures, and the anyons are topologically protected against the environmental noise. Finally, we estimated that the Ta$_4$Si$_4$Te$_4$ and Nb$_4$SiTe$_4$ chains have bandgap of 0.17 and 0.085 eV, respectively; this may potentially lead to a certain degree of topological protection even at room-temperature. Overall, the remaining challenge is that the parameter $\alpha$ does not have a topological protection, which makes the current proposal of quantum computation not completely topologically protected; similar problems also exist with the current Majorana-based qubits.

**Summary**

In conclusion, we present a one-dimensional topological model associated with an electric Zak phase and its relationship with the corresponding edge states. In our model, the nontrivial electric Zak phase of $\pi$ indicates the existence of a topological crystal insulator protected by a spinless mirror symmetry. The model involves only a single spin species, and exhibits a charge-fractionalization without many-body interactions. By material searching, we predict a series of materials, potentially within experimental reach, to probe the anomalous electric polarization. Based on the estimated size of the bandgaps, topological protection may be possible even at room temperatures. The topologically-protected boundary states can be employed for topological quantum computation, without superconductivity or external Zeenman fields. Systematic methods are presented for performing the braiding operations and encoding qubits. The qubits can be as small as few nanometers.

**Methods**

To explore the electronic structures and Zak phase of candidate materials, first-principles calculations are performed using Vienna ab initio simulation package (VASP) in the framework of the projector augmented waves. The cutoff energy is set as 600 eV, and a vacuum space larger than 20 Å is applied to decouple periodic interaction in two vertical directions to chains. The Perdew-Burke-Ernzerh of functional is adopted to describe the exchange-correlation interaction. All ions are relaxed until the force on every atom is smaller than 0.01 eV/ Å on a Γ-centered k-point mesh. The dependence of onsite Coulomb interaction *U* on band structure is checked, and *U* makes no significant effect on electric Zak phase and the main features of band structures. The ground-sate wave functions are determined by VASP, and then Wannier functions are obtained by Wannier90 packages.[43] At last, the boundary states of the real materials are determined by WannierTools package[44], in which the recursive Green's function is applied.

**Author contributions:** Z.G.S., J.S.Q, J.L., M.H.Y, J.B.Y conceived the calculation. Z.G.S. performed the coding and calculations. Z.G.S., M.H.Y, J.L., J.B.Y wrote the manuscript. All authors took part in the discussions and editing of the manuscript. J.Q. and Z.S. contributed equally to the work.

**Competing financial interests:**

The authors declare no competing financial interests.

**Corresponding authors:**

Correspondence and requests for materials should be addressed to Ju Li, Man-Hong Yung, Jinbo Yang. (email: Juli@mit.edu, yung@sustc.edu.cn or jbyang@pku.edu.cn).

**Acknowledgement:**

This work was supported by the National Key Research and Development Program of China (No. 2017YFA206303, 2016YFB0700901) MOST of China, National Natural Science Foundation of China (Grant Nos. 51731001, 51371009, 11674132, 11674005 and 11725415), National Basic Research Program of China (Grant No. 2016YFA0301004), and by the Key Research Program of the Chinese Academy of Sciences (Grant No. XPDPB08-4). JL acknowledges support by NSF DMR-1410636. We thank Zhengyuan Xue and Haizhou Lu for their discussions.

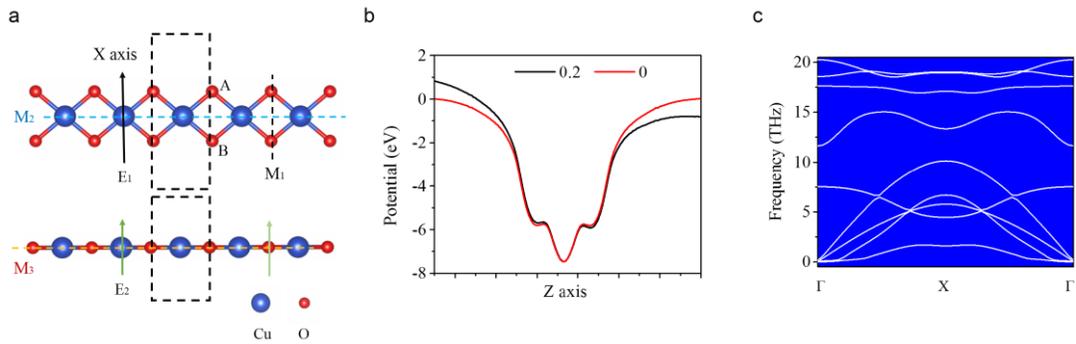

**Figure 1.** Geometrical structure and phonon dispersion of one-dimensional $CuO_2$. (a) Top (upper panel) and side (lower panel) views of one-dimensional $CuO_2$. The outline of the unit cell is sketched by black and dashed rectangles. The two distinguished O sites are labeled as A and B, and two orthogonal electric fields vertical to chain are labeled as $E_1$ and $E_2$. Mirror planes are denoted as $M_1$, $M_2$ and $M_3$, respectively. (b) Comparison between Hartree potential without and with an external electric field ($E_1$=0.2 V/Å). (c) Phonon dispersion calculated by DFT based on a supercell of $27 \times 1 \times 1$ single cells.

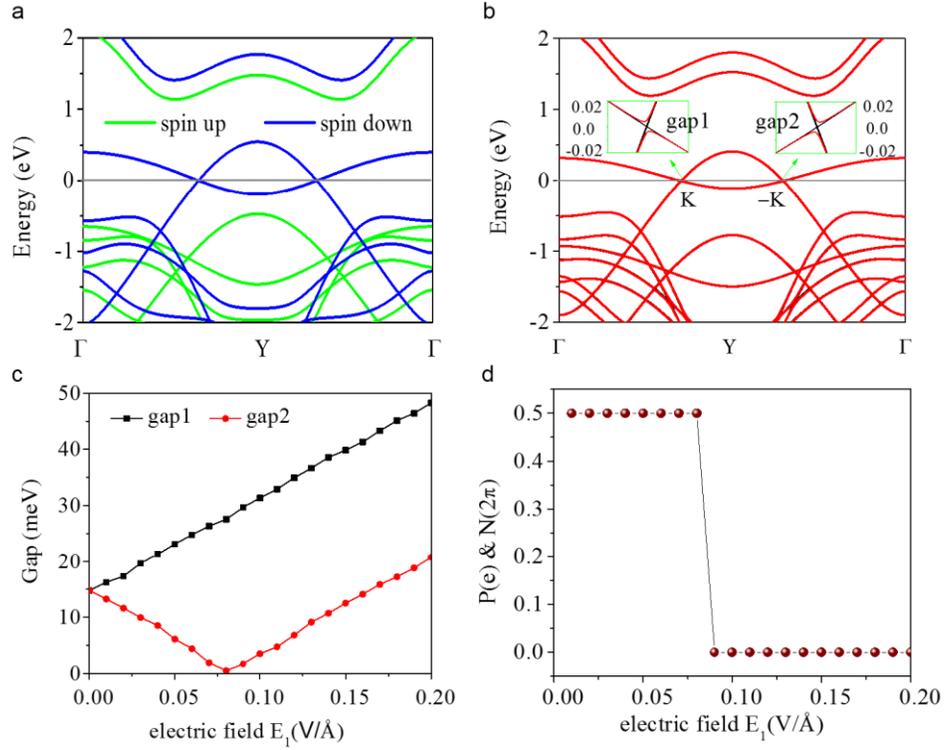

**Figure 2.** Electronic structure of the linear $CuO_2$ chain and its responses to electric fields. (a) Spin-resolved energy dispersion with quasi momentum. (b) Energy dispersion with quasi momentum including SOC. Red and black are the energy dispersions based on noncollinear spin with and without SOC, respectively. (c) Local bandgaps as a function of electric field $E_1$. (d) Electric polarization (P) and Zak phase winding number ($\mathbb{N}$) as a function of electric field $E_1$.

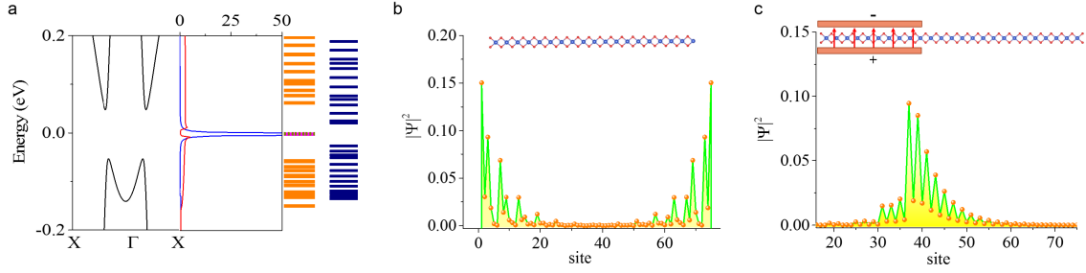

**Figure 3.** Electronic structure of edge states. (a) Bulk bands, density of states of semi-infinite and energy levels of finite strips. The orange and blue lines represent energy levels with and without SOC, respectively. (b) Electron density distribution of the zero-mode edge state localized at the edges. (c) Edge electron density distribution at the domain wall of trivial and nontrivial zones. The topologically trivial phase is realized by a large electric field. We amplify SOC and external electric field during plotting for the sake of clarity.

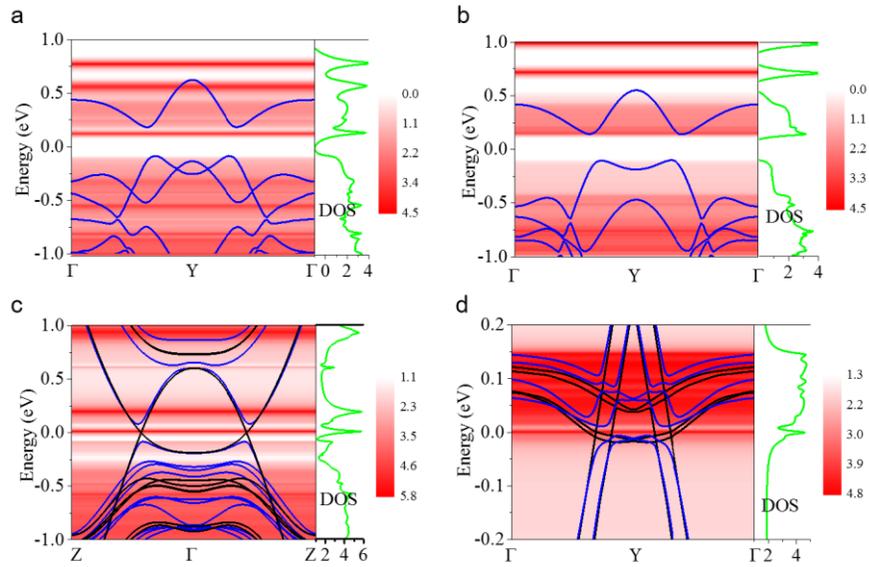

**Figure 4.** DFT calculated electronic structure of edge states. Blue and black curves are bulk band structures of different materials with and without spin orbit coupling, respectively. The green curves are the boundary density of sates. The red and white color represent the density of sates projected on the ends. (a) $CuO_2$ chain with an enhanced spin orbit coupling. (b) $CuO_2$ chain with broken inversion symmetry. (c-d) $Ta_4SiTe_4$ and $TiI_3$ chains.

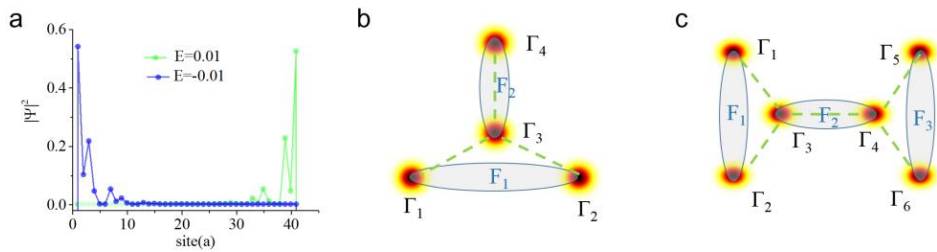

**Figure 5.** Braiding and encoding. (a) Electron density distribution of in-gap states under small differences of the onsite potential. (b) Configure of single qubit. (c) Configure of two-qubit setup.